\documentclass[journal=jacsat,manuscript=article]{achemso}

\usepackage{graphicx}% Include figure files
\usepackage{chemformula}
\usepackage{graphicx}
\usepackage{dcolumn}
\usepackage{bm}
\usepackage{float}
\usepackage{mathtools}   
\usepackage{algorithm}
\usepackage{algpseudocode}
\usepackage{mathptmx}
\usepackage{amsmath,amsthm,amssymb}
\usepackage{siunitx}
\usepackage{gensymb}
\usepackage[utf8]{inputenc}
\usepackage[T1]{fontenc}
\newcommand{\epsd}{\epsilon^\mathrm{ML}_\mathrm{d}} 
\newcommand{\epso}{\epsilon^\mathrm{ML}_\mathrm{o}}
\newcommand{\qmsbt}{QM7b-T}
\newcommand{\gdbt}{GDB-13-T}

\newcommand\footnoteref[1]{\protected@xdef\@thefnmark{\ref{#1}}\@footnotemark}

\author{Lixue Cheng}
\affiliation{Division of Chemistry and Chemical Engineering, California Institute of Technology, Pasadena, CA 91125, USA}
\author{Jiace Sun}
\affiliation{Division of Chemistry and Chemical Engineering, California Institute of Technology, Pasadena, CA 91125, USA}
\author{Thomas F. Miller III}
\email{tfm@caltech.edu.}
\affiliation{%
Division of Chemistry and Chemical Engineering, California Institute of Technology, Pasadena, CA 91125, USA%\\This line break forced% with \\
}%

\title{Accurate Molecular-Orbital-Based Machine Learning Energies via Unsupervised Clustering of Chemical Space}
\keywords{machine learning, electronic structure,  unsupervised clustering,
electron correlation, Gaussian processes}

\date{\today}% It is always \today, today,
\SectionNumbersOn 
\begin{document}
\begin{abstract}
We introduce an unsupervised clustering algorithm to improve training efficiency and accuracy in predicting energies using molecular-orbital-based machine learning (MOB-ML). 
This work determines clusters via the  Gaussian mixture model (GMM) in an entirely automatic manner and simplifies an earlier supervised clustering approach [J. Chem. Theory Comput., \textbf{15}, 6668 (2019)] by eliminating both the necessity for user-specified parameters and the training of an additional classifier. 
Unsupervised clustering results from GMM have the advantage of accurately reproducing chemically intuitive groupings of frontier molecular orbitals and having improved performance with an increasing number of training examples.
The resulting clusters from supervised or unsupervised clustering is further combined with scalable Gaussian process regression (GPR) or linear regression (LR) to learn molecular energies accurately by generating a local regression model in each cluster.
Among all four combinations of regressors and clustering methods, GMM combined with scalable exact Gaussian process regression (GMM/GPR) is the most efficient training protocol for MOB-ML.
The numerical tests of molecular energy learning on thermalized datasets of drug-like molecules demonstrate the improved accuracy, transferability, and learning efficiency of GMM/GPR over not only other training protocols for MOB-ML, i.e., supervised regression-clustering combined with GPR (RC/GPR) and GPR without clustering. GMM/GPR also provide the best molecular energy predictions compared with the ones from literature on the same benchmark datasets.
With a lower scaling, GMM/GPR has a 10.4-fold speedup in wall-clock training time compared with scalable exact GPR with a training size of 6500 \qmsbt{} molecules.
\end{abstract}

\maketitle

\section{Introduction}
\label{introduation}
Machine learning (ML) approaches have attracted considerable interest in the chemical sciences for a variety of applications, including molecular and material design, \cite{Gawehn2016, Popova2018, kearnes2016molecular, Mater2019, Kim2017, Ren2018, Butler2018, Sanchez-Lengeling2018, Mater2019} protein property prediction \cite{yang2019machine,casalino2020,gussow2020machine}, reaction mechanism discovery \cite{Wei2016, Raccuglia2016, Ulissi2017,gao2018using, Segler2018, Mater2019}, and analysis and classification tasks for new physical insights \cite{aarva2019understanding,zhang2020identifying, magdau2020}.
As an alternative to physics-based computations, ML has also shown promise for the prediction of molecular energies \cite{Smith2017, Smith2019, Lubbers,schutt2017quantum, wu2018moleculenet,Nguyen2018, Fujikake2018, Li2018,Zhang2018, Nandy2018, bogojeski2020quantum, glick2020ap,deephf,dick2020machine, christensen2020role}, intermolecular interactions \cite{glick2020ap, mezei2020noncovalent}, electron densities \cite{wu2018moleculenet, Grisafi2019, Pereira2018, Smith2019, fabrizio2019electron}, and linear response properties \cite{Ramakrishnan2015, Gastegger2017, Yao2018, Christensen2019, Ghosh2019, veit2020predicting}.
ML applications in the chemical sciences often rely on atom- or geometry-specific representations despite the increasing availability and feasibility of wavefunction-specific and deep-learning representations. \cite{Tuckerman, mcgibbon2017improving,Takuro2019,Townsend2019,deephf,dick2020machine,Welborn2018,Cheng2019,husch2020improved}.
Among these recent approaches, molecular-orbital-based machine learning (MOB-ML) \cite{Welborn2018, Cheng2019,cheng2019regression,husch2020improved,lee2020analytical} has been shown to exhibit excellent learning efficiency and transferability for the prediction of energies from post-Hartree-Fock wavefunction methods.

The previous versions of MOB-ML protocols \cite{Welborn2018, Cheng2019,husch2020improved}are limited to low training sizes due to the high computational cost of Gaussian process regression (GPR) training. A local regression with supervised clustering algorithm, termed as Regression-clustering (RC) algorithm (RC/GPR)\cite{cheng2019regression}, has been introduced in our prior work to reduce training costs and enable training on large datasets. RC/GPR clusters the training feature space with additional information from the label space of pair energies and classifies the test feature space by an additional classifier trained on feature space only. The learning efficiency of RC/GPR is mainly affected by the classification errors of the latter. 
Therefore, it is critical to adapt an efficient unsupervised clustering to bypass the introduction of this additional classification.
Husch et al. \cite{husch2020improved} propose an improved MOB feature, termed size-consistent feature, design by consistently ordering and numerically adjusting the features. The introduction of this improved MOB feature design not only enhances the prediction accuracy and transferability of MOB-ML, but also enables unsupervised clustering in chemical space. 

In this work, we apply a more accurate unsupervised clustering method, namely the Gaussian mixture model (GMM), to cluster systems in the organic chemical space using MOB features. Without any additional information from the label space, the resulting clusters from GMM agree with chemically intuitive groupings of molecular orbital (MO) types. To increase the learning efficiency for molecular energies, we construct the local regression models using a scalable GPR algorithm with exact GP inference but a lower scaling, i.e., alternative implementation of blackbox matrix-matrix multiplication (AltBBMM) algorithm, introduced in Ref. ~\citenum{sun2021molecular}. 
All the regression with clustering (supervised or unsupervised) methods offer exceptional efficiency and transferability for molecular energy learning. GPR with GMM clustering (GMM/GPR) is the most efficient training protocol for MOB-ML and delivers the best accuracy on \qmsbt{} and best transferability on \gdbt{} compared to other models in the literature. It also provides an over 10-fold wall-clock training time reductions compared to learning without clustering by AltBBMM.

\section{Theory}
\label{Theory}
\subsection{MOB-ML} 
Molecular energies can be expressed as a sum of Hartree-Fock (HF) energies and correlation energies ($E_c$). 
Utilizing Nesbet's theorem (Eq. \ref{Ecorr1}),\cite{Nesbet1958,SzaboNesbet} which states that $E_c$ for any post-Hartree-Fock wavefunction theory can be written as a sum over pair energies of occupied MOs, MOB-ML is a strategy that learns pair energies using features comprised of elements from the Fock, Coulomb, and exchange matrices.\cite{Welborn2018} Specifically,
\begin{equation}
\label{Ecorr1}
E_\textrm{c}=\sum^{\textrm{occ}}_{ij}\epsilon_{ij},
\end{equation}
where $\epsilon_{ij}$ is the pair energy corresponding to occupied MOs $i$ and $j$, and $\epsilon_{ij}$ is further expressed as a functional of the set of occupied and virtual MOs (Eq. \ref{Ecorr2}).
\begin{equation}
\label{Ecorr2}
\epsilon_{ij} = \epsilon\left[\{\phi_p\}^{ij} \right],
\end{equation}
MOB features are the unique elements of these matrices between $\phi_i$, $\phi_j$, and the set of virtual orbitals.
For maximum transferability between chemical systems, these properties are computed using localized MOs (LMOs)\cite{Welborn2018}. 
According to Eq. \ref{Ecorr2}, $\epsilon$ maps the HF MOs to the pair energies as a universal function, and $\epsilon$ can be approximated with two learned functions, $\epsd[\mathbf{f}_i]$ and $\epso[\mathbf{f}_{ii}]$, using feature vectors $\mathbf{f}_i$ (corresponding to $\mathbf{f}_{ii}$) and $\mathbf{f}_{ij}$ composed by MOB features, respectively (Eq. \ref{eq:diag_and_offdiag}).\cite{husch2020improved, Welborn2018, Cheng2019, cheng2019regression} In this study, we employ the same feature generation and sorting approach described in Ref. \citenum{husch2020improved}. 

\begin{equation}
    \label{eq:diag_and_offdiag}
    \epsilon_{ij} \approx
    \begin{cases}
        \epsd \left[\mathbf{f}_i\right] & \text{if $i=j$} \\
        \epso \left[\mathbf{f}_{ij}\right] & \text{if $i\ne j$}
    \end{cases}
\end{equation}
\subsection{Supervised and unsupervised clustering schemes for chemical spaces}

A straightforward application of Gaussian progress regression (GPR) with MOB features encounters a bottleneck due to computational demands since GPR introduces the complexity of $O(N^2)$ in memory and $O(N^3)$ in training cost. 
The property of local linearity for MOB features, which allows pair energies to be fitted as a linear function of MOB features within local clusters, has been investigated previously.\cite{cheng2019regression} Thanks to this property,
we proposed a comprehensive framework for local regression with clusters to further scale MOB-ML to the large data regime with lower training costs in Ref. \citenum{cheng2019regression}. Our previous work applied supervised clustering, i.e., regression-clustering (RC), to the training set and then performed GPR or linear regression (LR) as local regressors. A random forest classifier (RFC) was also trained to classify the test data. 

However, supervised clustering has its limitations. RC requires a predetermined number of clusters and an additional classifier \cite{cheng2019regression}. The performance of the supervised clustering scheme is also hindered by the classifier, which struggles to classify the results from RC due to the fact that the pair energy label information is only provided to RC but not RFC \cite{cheng2019regression}.  
Therefore, a more precise and efficient clustering and classification strategy is needed to enhance the performance of the entire framework. 

Improved MOB feature engineering results in a continuous MOB feature space \cite{husch2020improved}, and consequently enables the unsupervised clustering scheme for MOB-ML.
The points close in the feature space have similar chemical groupings, cluster identities, and label values, and therefore distance is an appropriate measure to cluster the MOB feature space. As a result, any distance-based clustering approach should perform well using the improved MOB features. K-means clustering is the simplest and fastest distance-based unsupervised clustering method, which can effectively cluster the MOB feature space and produce reliable regression results when used in conjunction with GPR. Unfortunately, the lack of an intrinsic probability measure and the unit ball assumption of k-means makes it not as accurate as other distance-based clustering methods, such as DBSCAN\cite{ester1996density}, OPTICS\cite{ankerst1999optics} and Gaussian mixture models (GMM). 

GMM can be treated as a generalized k-means method and is chosen for further investigation in this study. It assumes that all the $N$ data points belong to a mixture of a certain number of multivariate Gaussian distributions in the feature space with undetermined means and covariance, with each distribution representing a cluster. 
For a number of $K$ clusters (or Gaussian distributions) with $D$ feature dimensions, the cluster centers (or means of the distributions) $\{\mu_i \in \mathbb{R}^D, i=1, 2, ..., K\}$ and their corresponding covariance matrices $\{\Sigma_i \in \mathbb{R}^{D \times D}, i=1, 2, ..., K\}$ are solved by maximizing the likelihood $L$ using Expectation-Maximization (EM) algorithm.
The expectation, parameters, and clusters identities are computed and reassigned in the Expectation (E) stage, and the parameters to maximize likelihood are updated in the Maximization (M) stage. The two stages are repeated until reaching convergence.
For a test point, GMM can not only provide hard cluster assignments with the maximum posterior probability, but also enable soft clustering by computing the normalized posterior probability of a test point belonging to each cluster \cite{bishop2006pattern}. 
To make the GMM training completely automatic, we also perform model selection using the Bayesian information criterion (BIC) to determine the number of clusters used in GMM via scanning a reasonable series of candidate cluster sizes based on the training size $N$. BIC penalizes the likelihood increase due to including more clusters and more fitting parameters to avoid overfitting with respective to the number of clusters (Eq. \ref{eq:bic}),
 \begin{equation}
 \label{eq:bic}
      \text{BIC}=q\text{ln}(N)-2\text{ln}(L),
 \end{equation}
where $q$ is the number of parameters in the GMM model. 

\subsection{Local regression by alternative black-box matrix-matrix multiplication (AltBBMM) algorithm}
While the general framework of regression with clustering considerably improves the efficiency of MOB-ML\cite{cheng2019regression}, local regression with full GPR with a cubic time complexity remains the computational bottleneck for MOB-ML.
Recently, AltBBMM has been proposed to speed up and scale the GP training in MOB-ML for molecular energies with exact inferences. AltBBMM reduces the training time complexity to $O(N^2)$, enables the training on 1 million pair energies, or equivalently, 6500 \qmsbt{} molecules, and allows multi-GPU usage without sacrificing transferability across chemical systems of different molecular sizes. By applying AltBBMM as the local regressor, the efficiency of MOB-ML training can be significantly increased while incurring lower computational costs. The derivation and implementation details for AltBBMM are discussed in Ref. \citenum{sun2021molecular}.

\section{Computation details}
The performance of clustering and subsequent local regression approaches are evaluated on \qmsbt{} and \gdbt{} benchmark systems, which comprise molecules with at most seven and only thirteen heavy atoms (C, O, N, S, and Cl), respectively.
Each molecule in \qmsbt{} and \gdbt{} has seven and six conformers, respectively, and only one conformer of each randomly selected \qmsbt{} molecule is used for training.
The features are computed at HF/cc-pVTZ level with the Boys–Foster localization scheme\cite{Boys1960,foster_canonical_1960} using \textsc{entos qcore} \cite{manby2019entos} and reference MP2\cite{Moller1934,LMP2} pair energy labels with cc-pVTZ basis set \cite{Dunning1989} are generated from Molpro 2018.0\cite{MOLPRO}. All the features, selected features and reference pair energies employed in the current work are identical to those reported in Ref. \citenum{husch2020improved}. 

\subsection{Supervised Clustering with MOB features}
\label{sec:supervised}
RC can cluster the organic chemical space represented by \qmsbt{} and \gdbt{}\cite{cheng2019regression} by maximizing the local linearity of the MOB feature space. 
On the MOB feature space, we apply the same standard RC protocol introduced in Ref. \citenum{cheng2019regression} using k-means cluster initialization \cite{cheng2019regression} and ordinary least square linear regressions (LR) implemented using \textsc{CuPy}\cite{cupy_learningsys2017}. The RC step is fully converged (zero training MAE change between two iterations) to obtain the training clusters. Random forest classification (RFC) with 200 balanced-tress implemented in \textsc{Scikit-learn}\cite{cheng2019regression} is performed to classify the test data. To reduce the cost of training RFC and local regressors for the off-diagonal clusters with a large number of pairs, we adapt the same capping strategy illustrated in Ref.\citenum{cheng2019regression} with a capping size of 10,000 for each training off-diagonal cluster during the training over 2000 \qmsbt{} molecules. There is no capping applied to all diagonal and off-diagonal pairs with training sets of fewer than 2000 molecules.  

\subsection{Unsupervised clustering with MOB features}
Following the implementation in \textsc{Scikit-learn}, we reimplement GMM to enable multi-GPU usage using \textsc{CuPy}, which is initialized by k-means clustering and constructed with a full covariance matrix. The objective function of GMM is to maximize the likelihood, which is solved iteratively by the EM algorithm. A regularization of 1e-6 is added to its diagonal terms to ensure the positive definiteness of the covariance matrix of GMM.

The number of clusters $K_{best}$ used in GMM is automatically detected by scanning a series of reasonable cluster sizes and finding the GMM model with the most negative BIC score. According to the previous study in Ref. \citenum{cheng2019regression}, the optimal numbers of clusters for the diagonal and off-diagonal models of 1000 training molecules in RC are 20 and 70, respectively. The scanning series of possible $K$ are $\{5i| i = 1,2,...,10\}$ and $\{5i| i=7,8,...,32\}$ for diagonal and off-diagonal pairs, respectively. Empirical equations for estimating the scanning range of the number of clusters are also presented. We note that this auto-determination procedure is completely unsupervised and does not require any cross-validation from regression. 

A hard clustering from GMM assigns the test point to the cluster with the highest probability, and a soft clustering from GMM provides probabilities of the point belonging to each possible cluster. Only a few pairs (under 10\%) in \qmsbt{} can have a second most probable cluster with a probability over 1e-4 (Table S3). More details about soft clustering are described in the SI. The current work presents and analyzes the results from hard clustering without specifying any parameters. To demonstrate the smoothness and continuity of GMM clusters constructed using MOB features on the chemical space, the Euclidean distance between the feature vector of each diagonal pair and the corresponding hard cluster center $\mu_i$ is computed and analyzed.

\subsection{MO type determinations}
To compare the cluster compositions and the MO compositions in \qmsbt{} and \gdbt{}, we also apply an algorithm to determine MO types represented by atomic connectivity and bond order for closed-shell molecules following the octet rule. This procedure requires the coordinates of atoms and the centroids of MOs computed using HF information. More details and the pseudo-code of the MO detection algorithm are included in the SI (Algorithm 1 in SI).

\subsection{Regression within local clusters}
Regressions by GPR or LR on top of RC or GMM clustering are used to predict molecular energies. For LR, we use the ordinary least square linear regression with no regularization for diagonal and off-diagonal pairs. 
To reduce the training cost of local GPRs, AltBBMM \cite{sun2021molecular} is performed with Mat\'ern 5/2 kernel with white noise regularization of 1e-5 for both diagonal and off-diagonal pair energies. For the clusters with training points fewer than 10,000, GPR models are directly obtained by minimizing the negative log marginal likelihood objective with the BFGS algorithm until full convergence. For the clusters with more than 10,000 training points, the variance and lengthscale are first optimized using randomly selected 10,000 training points within the cluster, and the Woodbury vector \cite{rasmussen2006} is further solved by the block conjugate gradient method with preconditioner sizes of 10,000 and block sizes of 50.

In order to improve the accuracy and reduce uncertainty, without specifications, the predicted energies are reported as the averages of ten independent runs for all MOB-ML with clustering protocols. We abbreviate the RC then RFC classification and GPR regression as RC/GPR since no other classifier is used with RC. Similarly, RC then RFC classification and LR regression, GMM clustering with GPR regression, and GMM clustering with LR regression are abbreviated as RC/LR, GMM/GPR, and GMM/LR, respectively. The entire workflow on the general framework of MOB-ML with clustering is also introduced in Ref. \citenum{cheng2019regression}.

\section{Results and discussions}
\subsection{Number of clusters detected in GMM}
\label{sec:cluster_num}
\begin{figure}[htbp]
\includegraphics[width=0.65\columnwidth]{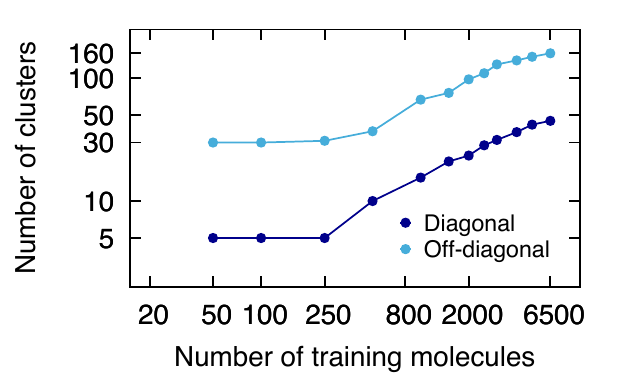}
\caption{Numbers of clusters in GMMs for diagonal and off-diagonal pairs detected by BIC scores. The average number of clusters over ten runs is plotted versus the number of training molecules in \qmsbt{} on a logarithm scale. 
}
\label{figure:cluster_num}
\end{figure}

Rather than predetermining the number of clusters through pilot experiments \cite{cheng2019regression}, GMM automatically selects the most suitable model among the ones with different cluster numbers by finding the lowest BIC score, which prevents overfitting due to a large number of clusters and is faithful to the intrinsic feature space structure in the training set \cite{findley1991counterexamples}. 
Figure \ref{figure:cluster_num} depicts the optimal number of clusters determined by BIC scores as a function of the number of training \qmsbt{} molecules. The numbers of diagonal and off-diagonal clusters are roughly proportional to the training sizes in a logarithm scale if the training set is larger than 250 molecules, and the best number of clusters can be estimated as functions of the number of training molecules $N_{mol}$ from this set of results as $K_{d}$ and $K_{o}$ for diagonal and off-diagonal pairs, respectively.
\begin{align}
    %K_d=\text{exp}(\text{ln}(N_{mol}^0.579)-1.219),\\
    K_d&=0.296\, N_{mol}^{0.579},\\
    %K_o=\text{exp}(\text{ln}(N_{mol}^0.502)+0.750)
    K_o&=2.117\, N_{mol}^{0.502}
\end{align}

These two empirical equations serve as estimation functions to avoid searching an excessive amount of candidate clustering numbers. For future multi-molecule dataset trainings, it is sufficient to construct the scanning region of possible $K$ values as [$K_{est}-10$, $K_{est}-5$, $K_{est}$, $K_{est}+5$, $K_{est}+10$], where $K_{est}$ is the five multiple closest to the estimated value computed by the above empirical equations.

\subsection{Unsupervised clustering organic chemical space}
\subsubsection{Chemically intuitive clusters from unsupervised clustering}
\label{sec:qm7b_gmm}
\begin{figure*}[hbtp]
\includegraphics[width=0.99\columnwidth]{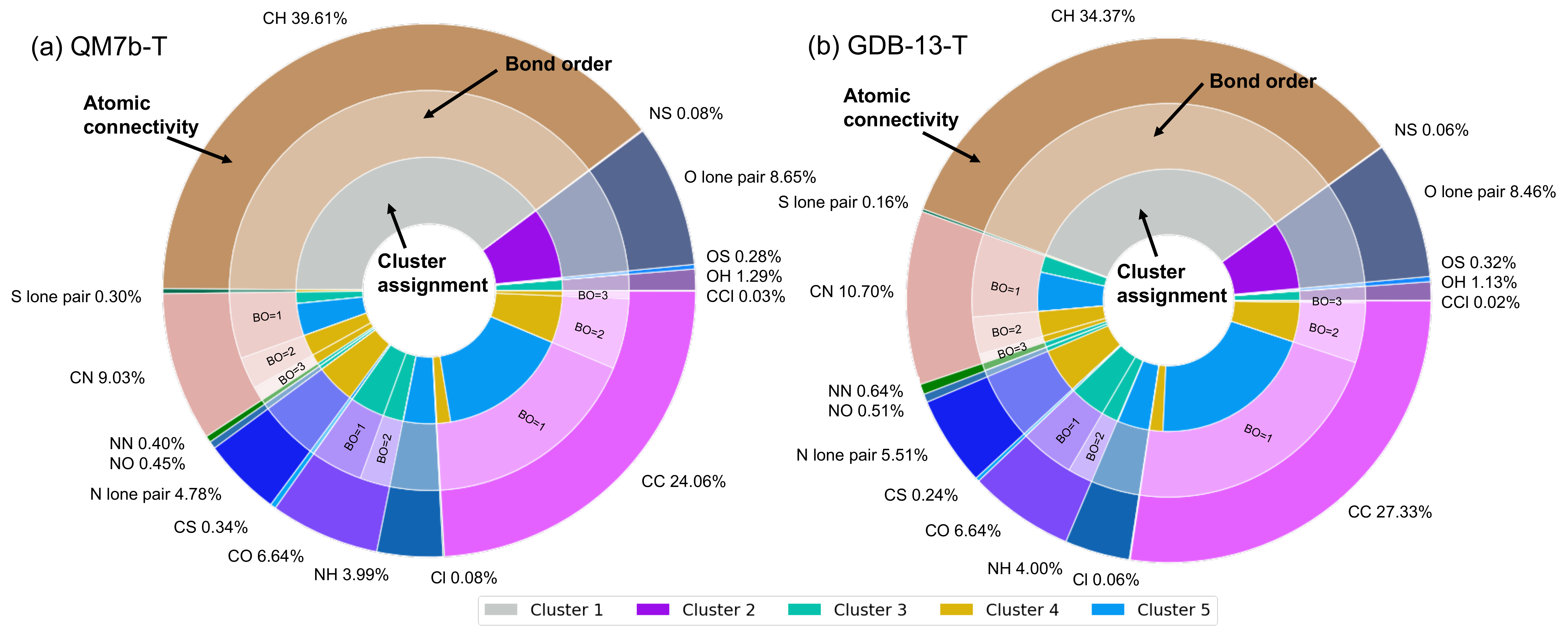}
\caption{MO types and cluster compositions of (a) \qmsbt{} and (b) \gdbt{} predicted by GMM model trained on diagonal features of 250 \qmsbt{} molecules (250 GMM model). The layers from outer to inner are the atomic connectivity of MOs, the bond orders (BO) of MOs, and the GMM classification results, respectively. 
The abundance of each type of atomic connectivity in each dataset is labeled.
The BOs are only marked for cases where one type of atomic connection has more than one possible bond order. If the two atoms of a MO can only form a single bond or the MO is a lone pair, the bond order is not listed in the figure.
}
\label{figure:qm7b_clustering}
\end{figure*}

A specific MO can be one-to-one represented by its diagonal feature space, and thus all the MO analyses are conducted with the clustering results from GMMs trained on diagonal features using different numbers of \qmsbt{} molecules. The GMM clustering results and MO types are categorized by multiple pie charts layer by layer for \qmsbt{} and \gdbt{} datasets in Fig. \ref{figure:qm7b_clustering}.
The first (outermost) layer depicts the atomic connectivity of a MO, which is further classified by the bond order in the second (intermediate) layer. The third (innermost) layer illustrates the classification results for each type of MO obtained from the diagonal GMM model trained on 250 \qmsbt{} molecules.

MOB-ML has been shown to be transferable in supervised clustering and regression tasks by creating an interpolation to weak extrapolation tasks between different chemical systems using the MOB representation. \cite{Welborn2018, Cheng2019, cheng2019regression, husch2020improved} When the first and second layers of two sets of pie charts in Fig. \ref{figure:qm7b_clustering}a and b are compared, it becomes clear that \qmsbt{} and \gdbt{} share the same categories of MOs with slightly different abundances. The C-H MO is the most prevalent MO type in \qmsbt{}, and its popularity declines as the popularity of other less-trained MOs increases in \gdbt{}. This discovery implies that \qmsbt{} has a majority of the information necessary to predict the properties of molecules in \gdbt{} with any MO-based representation and any transferable machine learning approach.
The almost identical grouping patterns in the third layers of Fig. \ref{figure:qm7b_clustering}a and b suggest that unsupervised clustering via GMM is transferable as expected. 
In addition, the cluster assignments match the chemically intuitive groupings for both \qmsbt{} and \gdbt{} (Fig.\ref{figure:qm7b_clustering}a inner layer). Each type of MO is clustered into one type of cluster except for the C-C single bond, while most clusters contain more than one type of MOs. For example, all CC double bonds are clustered into Cluster 2, but Cluster 2 contains C-C single, double and triple bonds, C-N double and triple bonds, and lone pairs on the N atom. More training points are required to capture finer clustering patterns in the chemical space using GMM.

We note that while clustering based on MO types is theoretically feasible, it is not practical as a general approach in MOB-ML to predict molecular properties. 
It is difficult to intuitively define the types of MOs within chemical systems with complicated electronic structures, such as transition states. On the other hand, MOB features can easily represent these MOs.
As demonstrated in the first layers in both Fig. \ref{figure:qm7b_clustering}a and b, the organic chemical space is biased heavily towards C-H and C-C MOs significantly. To avoid overly small clusters and achieve accurate local regression models, a careful design of training sets is also required by including various MO types for clustering based on MO types. In addition, the local regression models with clusters based on MO types cannot predict the properties of a new type of MO without explicitly including it in the training set. Meanwhile, a GMM model trained with MOB features could still classify this MO into a suitable group that has the smallest feature space distances to the MO.

\subsubsection{Resolutions of GMM clustering with different training sizes}
\label{sec:split}

\begin{figure*}[htbp]
\includegraphics[width=0.9\columnwidth]{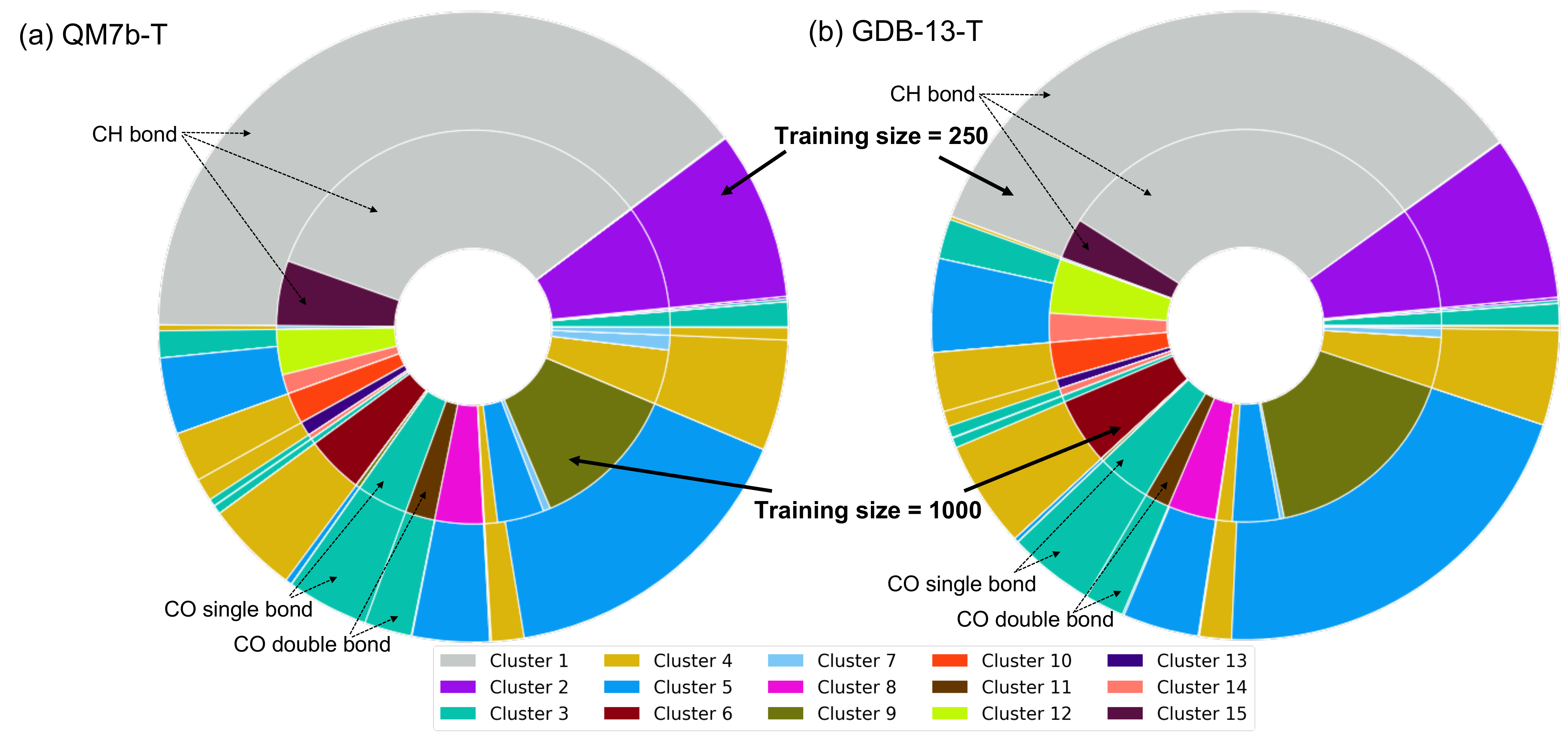}
\caption{Cluster assignments of (a) \qmsbt{} and (b) \gdbt{} predicted by the diagonal GMM models trained on 250 (250 GMM model) and 1000 (1000 GMM model) \qmsbt{} molecules. The outer layers display the same clustering results as the most inner layers in Fig. \ref{figure:qm7b_clustering} predicted by the GMM model trained on 250 molecules with five detected clusters. The inner layers show the clustering results predicted by the GMM model trained on 1000 molecules with 15 detected clusters. In both panels, the clusters in the inner layers further split up the ones in the outer layers. The MO identities of example clusters analyzed in the main text are labeled in the figure as well.
}
\label{figure:cluster_split}
\end{figure*}

As the number of training pairs increases, the number of clusters recognized by GMM for diagonal feature space increases from 5 at 250 training molecules to 15 at 1000 training molecules (Fig. \ref{figure:cluster_num}). Figure \ref{figure:cluster_split}a and b compares the clustering patterns predicted by GMMs trained on different training sizes for \qmsbt{} and \gdbt{}, respectively. In both panels, the layers show the cluster compositions determined by the GMM trained on 250 molecules (250 GMM model) in the outer layer and 1000 molecules (1000 GMM model) in the inner layer.
Training on more molecules not only provides more diverse chemical environments for the same type of MOs, but also aids in the resolution of the local structures in the MOB feature space. The MO types with high abundance in \qmsbt{} and \gdbt{} could be split into multiple clusters. For instance, the one cluster for CH single bond trained on 250 molecules is split into two clusters trained on 1000 molecules. In addition, the MO types with low abundances in \qmsbt{} and \gdbt{} could be resolved with more training data, rather than mixed into one cluster. For example, C-O single bonds and C-O double bonds are classified into two clusters by the 1000 GMM model instead of one cluster by the 250 GMM model.

\subsection{Molecular energy learning by regression with clustering}
\label{sec:correlation}
\begin{figure*}[bhtp]
\includegraphics[width=0.99\columnwidth]{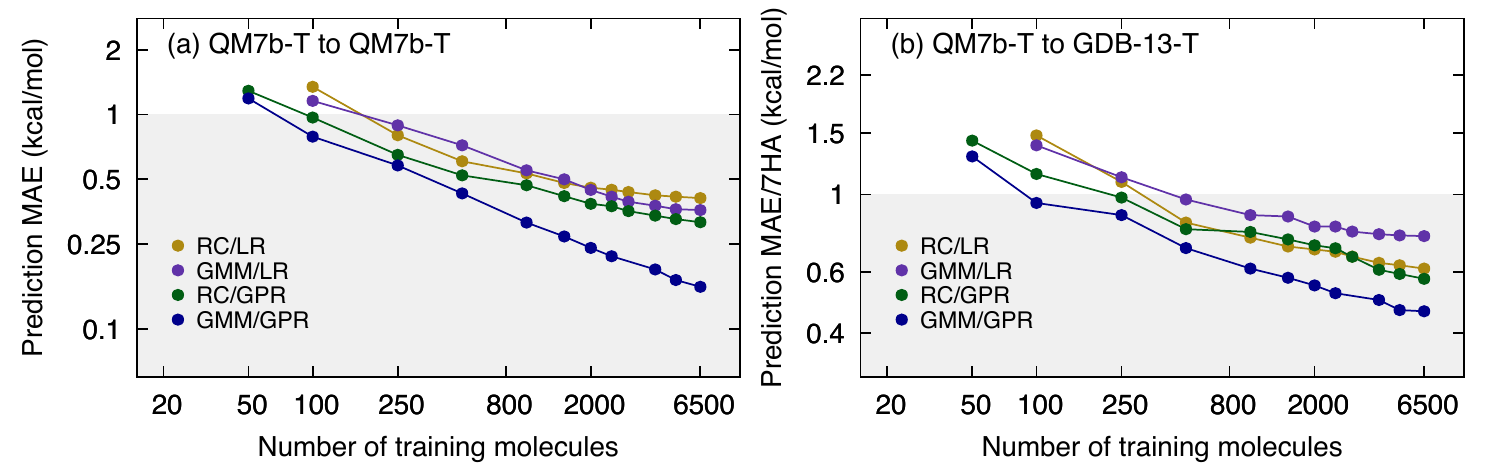}
\caption{Learning curves for MP2/cc-pVTZ  energy predictions with different clustering methods trained on \qmsbt{} and applied to (a) \qmsbt{} and (b) \gdbt{}. The models are the same ones trained on QM7b-T for (a) and (b). The prediction performance is reported in terms of MAEs and MAE per 7 heavy atoms (7HA) for (a) and (b), respectively, by averaging over ten runs.  All the data are plotted on a logarithm scale, and the shaded areas correspond to an MAE/7HA of 1 kcal/mol. 
}
\label{figure:qm7b_corr_thiswork}
\end{figure*}
We now present the results of predicting molecular energies utilizing GPR or LR on top of supervised or unsupervised clustering methods in MOB-ML. The RC/LR and GMM/LR training results on 50 molecules are omitted due to the instability of local LR models with 50 training molecules. The prediction accuracy is assessed by mean absolute error (MAE) of total energies predicted by each MOB-ML model on test sets and plotted as a function of the number of training molecules on a logarithm scale (``learning curve"\cite{learning_curves}) in Fig. \ref{figure:qm7b_corr_thiswork}. The test sets consist of all remaining \qmsbt{} thermalized geometries not included in the training sets in Fig.\ref{figure:qm7b_corr_thiswork}a and all the \gdbt{} thermalized geometries in Fig.\ref{figure:qm7b_corr_thiswork}b. All the test errors for different training protocols with different training sizes are reported in Supporting information Table S1 and Table S2.

Among all four training protocols, GMM/GPR provides the best learning accuracy on \qmsbt{} and transferability on \gdbt{}. By training on 6500 molecules, GMM/GPR can achieve an MAE of 0.157 kcal/mol and an MAE/7HA of 0.462 kcal/mol for \qmsbt{} and \gdbt{}, respectively. 
The performances of all three other approaches are similar on \qmsbt{}, but RC/GPR and RC/LR have slightly better performance on \gdbt{} than GMM/LR. For the models clustered by RC, LR provides similar accuracy and transferability compared with GPR since RC maximizes the local linearity for each local cluster. The accuracy loss due to non-linearity of local regression is more significant with GMM clustering with an MAE of 0.202 kcal/mol and an MAE/7HA of 0.298 kcal/mol for \qmsbt{} and \gdbt{}, respectively, training on 6500 molecules. Although GMM/LR is not as accurate as GMM/GPR, the reasonably accurate predictions from GMM/LR for both \qmsbt{} and \gdbt{} infer that GMM still can capture local linearity to some degree, despite the fact that GMM is not trained to maximize local linearity.
In comparison to GMM/GPR, the learning efficiency of RC/GPR is harmed by the classification errors from RFC for test points\citenum{cheng2019regression}, and hence RC/GPR provides twice as large errors for \qmsbt{} and 0.111 kcal/mol worse MAE/7HA for \gdbt{}.

\begin{figure*}[hbpt]
\includegraphics[width=0.95\columnwidth]{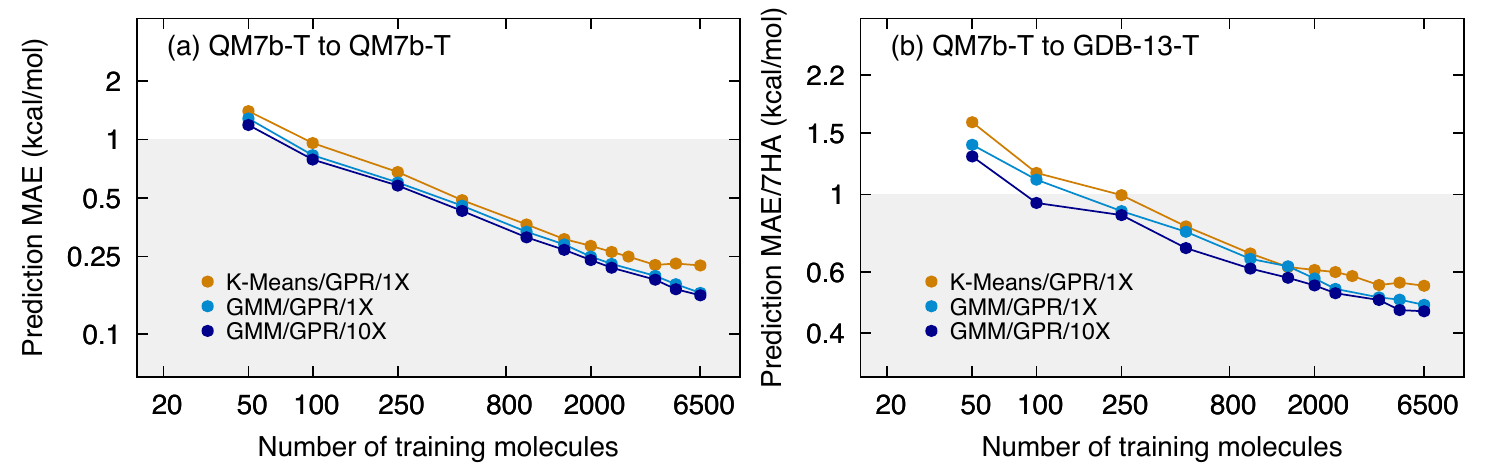}
\caption{Learning curves for MP2/cc-pVTZ energy predictions with different regression with unsupervised clustering methods trained on \qmsbt{} and applied to (a) \qmsbt{} and (b) \gdbt{}. The results of GMM/GPR/10X are the same as the ones of GMM/GPR in Fig. ~\ref{figure:qm7b_corr_thiswork} and plotted for comparison. All the data are plotted on a logarithmic scale, and the shaded areas correspond to an MAE/7HA of 1 kcal/mol. 
}
\label{figure:kmeans}
\end{figure*}
With the same clustering method, GPR is a more accurate local regressor compared with LR and generally offers superior accuracy across all the training sizes. GMM/LR has 1.47 to 2.28 times higher MAEs than GMM/GPR, and RC/GPR also marginally outperforms RC/LR. The chemical accuracy of 1 kcal/mol for test \qmsbt{} molecules can be reached by training on 100 and 250 training molecules using GPR and LR local regressors, respectively in Fig. \ref{figure:qm7b_corr_thiswork}a. In addition, GMM/GPR only requires 100 training molecules to reach the chemical accuracy for \gdbt{}. 

\subsection{Performances of alternative settings of regression with unsupervised clustering}

In this section, we show the results of some alternative settings of regression with unsupervised clustering. We compare the accuracy changes with and without averaging over ten runs and replace the GMM clustering with k-means to show the generality of the local GP with unsupervised learning.
Figure ~\ref{figure:kmeans} the learning curves of molecular energies obtained from different regression with unsupervised clustering learning protocols are plotted and compared with our standard GMM/GPR learning protocol. Although averaging over ten independent runs offers slightly better prediction accuracy for GMM/GPR, the results from single run GMM/GPR (GMM/GPR/1X) only have minor accuracy loss. The results obtained from a single run of GMM/GPR (GMM/GPR/1X) are slightly worse than those averaging over ten independent runs (GMM/GPR/10X). The best accuracy of GMM/GPR/1X is MAE=0.162 kcal/mol, which is only 3\% worse than the one of GMM/GPR/1X for QM7b-T. The transferability of predicting GDB-13-T also remains nearly unchanged. 
However, RC/GPR/1X is over 10\% worse than the corresponding RC/GPR/10X in Ref. \cite{cheng2019regression}. This observation suggests that GMM is not only more accurate but also more stable compared with RC.  

This general framework of regression with unsupervised clustering does not restrict the usage of other unsupervised clustering algorithms, such as k-means and density-based spatial clustering of applications with noise (DBSCAN). Here, we include the learning curves of k-means with GPR as regressors (k-means/GPR/1X) in Fig. \ref{figure:kmeans}. The number of clusters in k-means is auto-detected by the Davies-Bouldin index\cite{davies1979cluster}, which compares the distance between clusters with the size of the clusters themselves\cite{scikit-learn} since BIC could not be computed without Bayesian inferences. K-means is found to be a reasonably good choice for clustering and classification. The prediction accuracies of k-means/GPR/1X are only at most 36.57\% and 10.30\% worse for \qmsbt{} and \gdbt{} than the ones of GMM/GPR/1X, respectively, which indicates the potential success of using any other unsupervised clustering algorithms in MOB-ML.

\subsection{Comparison with molecular energy learning results from literature}
\label{sec:lit}
\begin{figure*}[hbtp]
\includegraphics[width=0.95\columnwidth]{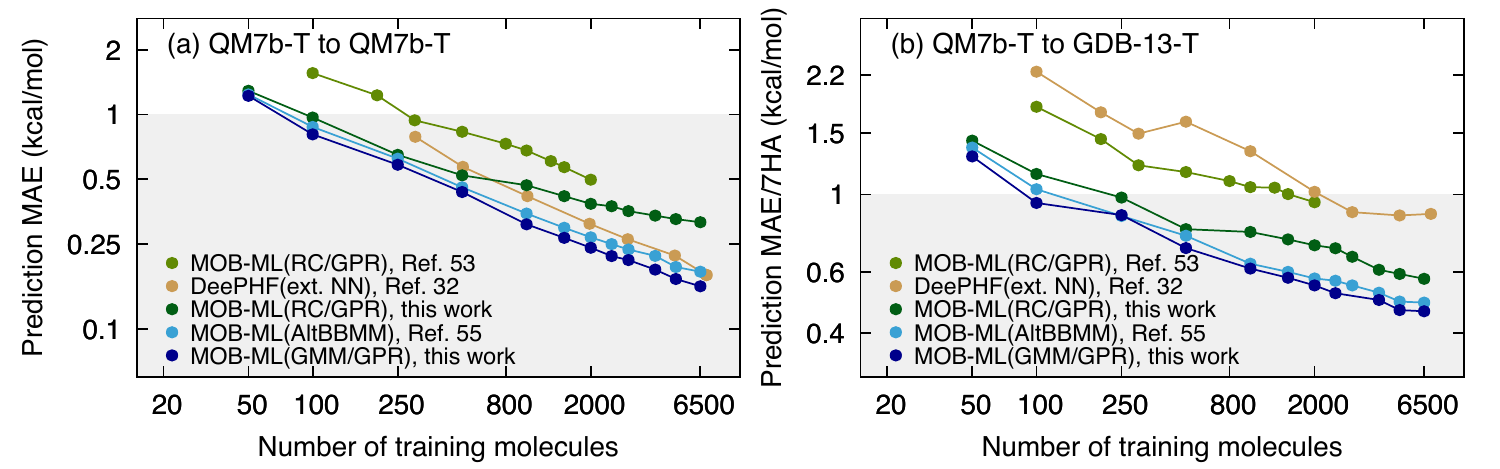}
\caption{Accuracy comparison between
different ML methods trained on \qmsbt{} and tested on (a) \qmsbt{} and (b) \gdbt{}. The learning curves of RC/GPR and GMM/GPR are the same ones shown in Fig. \ref{figure:qm7b_corr_thiswork}. Results from RC/GPR in Ref. \citenum{cheng2019regression} were trained on non-size consistent features and therefore different from the ones obtained from RC/GPR in this work. In addition, MOB-ML regressed with AltBBMM (MOB-ML (AltBBMM)) \cite{sun2021molecular} and DeePHF (ext. NN) \cite{deephf} are also plotted for comparison. All the data are digitally extracted from the corresponding studies and plotted on a logarithm scale. The shaded area corresponds to the chemical accuracy of 1 kcal/mol. 
}
\label{figure:qm7b_corr_lit}
\end{figure*}

In Fig.\ref{figure:qm7b_corr_lit}, the learning curves of RC/GPR and GMM/GPR in this study are further compared  to those of state-of-the-art methods in the literature trained on randomly selected \qmsbt{} molecules, including MOB-ML regressed with RC/GPR using outdated MOB features from Ref. \citenum{cheng2019regression} (MOB-ML (RC/GPR)), DeePHF trained with an NN regressor \cite{deephf} (DeePHF (ext. NN)), and MOB-ML regressed with AltBBMM (MOB-ML (AltBBMM)) \cite{sun2021molecular}.

The introduction of the most recent improved MOB features \cite{husch2020improved} considerably enhances the accuracy of MOB-ML, and therefore RC/GPR from this work is more accurate than the literature RC/GPR with outdated features. Training on the best available MOB features leads to over 30 \% accuracy improvements with RC/GPR on both \qmsbt{} and \gdbt{} test molecules. This observation suggests that better feature engineering not only can improve the accuracy for GPR without clustering \cite{husch2020improved}, but also can enhance the efficiency of regression with clusters. GMM/GPR achieves slightly higher prediction accuracy than AltBBMM without clustering when using the same MOB feature design, indicating that an additional GMM clustering step prior to regression benefits the entire training process by replacing the global regression model with more accurate local regression models.

As another machine learning framework to predict molecular energies at HF cost, DeePHF \cite{deephf} also achieves accurate predictions for \qmsbt{}, while its transferability on \gdbt{} is less than MOB-ML \cite{husch2020improved}.
GMM/GPR training on 6500 molecules (MAE=0.157 kcal/mol) outperforms the best DeePHF model training on 7000 molecules (MAE=0.159 kcal/mol) on the total energies of \qmsbt{} test molecules. Without sacrificing transferability on \gdbt{}, the best model from GMM/GPR can achieve half of the error from DeePHF on \gdbt{} and become the most accurate model for molecular energies in \gdbt{}.

\subsection{Efficient learning by local AltBBMM with GMM clustering}
\begin{figure}[htbp]
\includegraphics[width=0.7\columnwidth]{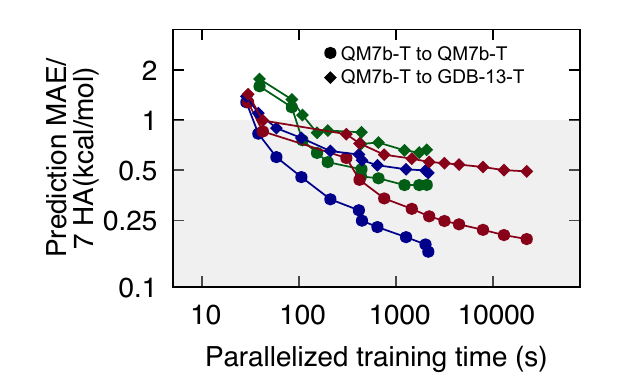}
\caption{Accuracy and training costs of MP2/cc-pVTZ energy using single-run GPR with RC and GMM clustering (RC/GPR/single-run, green; GMM/GPR/single-run, blue) and AltBBMM without clustering (red, Ref.~\citenum{sun2021molecular}). Prediction MAEs of test \qmsbt{} (circles) and \gdbt{} (diamonds) from single runs are plotted as a function of wall-clock training time with parallelization on 8 NVIDIA Tesla V100 GPUs on a log scale.
The models are the same as the ones reported in Fig. \ref{figure:qm7b_corr_thiswork} and the corresponding training sizes of \qmsbt{} are labeled in the figure.  
The shaded areas correspond to an MAE/7HA of 1 kcal/mol. 
}
\label{figure:total_timing}
\end{figure}

To have a fair comparison, we report the timings of single-run regression with clustering models and compared with the ones from AltBBMM models in this section. 
Figure \ref{figure:total_timing} plots the test MAEs of \qmsbt{} and \gdbt{} from single-run models as a function of parallelized training time on 8 NVIDIA Tesla V100-SXM2-32GB GPUs for three most accurate MOB-ML training protocols. 
GMM/GPR provides slightly improved accuracy and transferability compared to direct regression by AltBBMM without clustering and significantly reduces the training time of MOB-ML by 10.4 folds with 6500 training molecules. As the most cost-efficient and accurate training protocol for MOB-ML, a single run of GMM/GPR only requires 2170.4s wall-clock time to train the best model with 6500 molecules, while AltBBMM needs 22486.4s to reach a similar accuracy.

We note that the computational costs of GMM and local AltBBMM in GMM/GPR are comparable and lower than AltBBMM without clustering. The complexity analysis is as follows.
The training complexity of GMM of each EM iteration is $O(NK)$ with a fixed number of features\cite{pinto2015fast}, where $N$ is the number of training points and scales linearly with $N_{mol}$, and $K$ is the number of clusters.
Local AltBBMM has a training complexity of $O(KN_{loc}^2)$, where $N_{loc}$ is the number of training points in each local cluster\cite{rasmussen2006}. Since $N_{loc}$ roughly scales as $O(N/K)$, the complexity of local AltBBMM in GMM/GPR can be approximated as $O(N^2/K)$. Therefore, GMM becomes the computational bottleneck in GMM/GPR when $K$ grows faster than $N^{0.5}$; otherwise, local AltBBMM is more expensive than GMM.
As discussed in Sec. \ref{sec:cluster_num}, the optimal $K_d$ and $K_o$ for \qmsbt{} are fitted as functions of $N$ with an approximate scaling of $O(N^{0.579})$ and $O(N^{0.502})$, respectively. GMM and local AltBBMM share similar computational costs in this case, and the overall complexity of GMM/GPR using local AltBBMM is around $O(N^{1.58})$, which is lower than AltBBMM without clustering. By using this set of complexity analysis, the training size of MOB-ML is no longer limited by  GP, and MOB-ML is able to train more than 100 million pair energies (equivalent to 100,000 molecular energies).

\section{Conclusion} 
We extend our previous work on supervised clustering to unsupervised clustering on the organic chemical space with the improved MOB features. An accurate, efficient, and transferable regression with clustering scheme is also introduced to learn the molecular energies of \qmsbt{} and \gdbt{}. Without specifying the number of clusters, unsupervised clustering via Gaussian mixture model (GMM) is automatic without human interference and able to cluster the organic chemical space represented by \qmsbt{} and \gdbt{} in ways consistent with the chemically intuitive groupings of MO types. The finer grouping patterns of MOB feature space are captured as the amount of training data increases, and the resulting clusters are gradually separated following chemical intuition.
As the most efficient training protocol for MOB-ML, GMM/GPR surpasses RC/GPR and AltBBM without clustering in prediction accuracy and transferability with a training cost at a tenth of the one of AltBBBM without clustering.
GMM/GPR not only reaches the chemical accuracy for \qmsbt{} and \gdbt{} by only training on 100 \qmsbt{} molecules, but also offers superior performance to all other state-of-the-art ML methods in literature with an MAE of 0.157 kcal/mol for \qmsbt{} and an MAE/7HA of 0.462 kcal/mol for \gdbt{}. We finally illustrate that the overall complexity of GMM/GPR is lower than AltBBMM without clustering and local AltBBMM regression is no longer the computational bottleneck in GMM/GPR. 
As a future direction, it is promising to apply GMM/GPR to even larger datasets with more diverse chemistry due to its low computational complexity.
The unsupervised nature of GMM also opens an avenue to regress other molecular properties with MOB features by GMM/GPR.

\begin{acknowledgement}
We thank Dr. Tamara Husch for the guidance on the improved feature generation protocol, and Vignesh Bhethanabotla for his help to improve the quality of this manuscript. TFM acknowledges support from the US Army Research Laboratory (W911NF-12-2-0023), the US Department of Energy (DE-SC0019390), the Caltech DeLogi Fund, and the Camille and Henry Dreyfus Foundation (Award ML-20-196). Computational resources were provided by the National Energy Research Scientific Computing Center (NERSC), a DOE Office of Science User Facility supported by the DOE Office of Science under contract DE-AC02-05CH11231.
\end{acknowledgement}

\section*{Supporting Information}
Detailed descriptions of the algorithm of MO type detection (Algorithm S1) are included in the supporting information.
Table S1 and S2 in the supporting information provide the numerical data for the Fig. \ref{figure:qm7b_corr_thiswork} in the main text. 
We include the wall-clock time comparison of RC and GMM on clustering step only in Fig. S1,  learning curves of soft clustering from GMM combined with LR regressor in Fig. S2, and Table S3 shows the percentage of pairs whose predicted energies are affected slightly by soft clustering. 
\bibliography{main}% Produces the bibliography via BibTeX.
\end{document}